\documentclass{aastex61}

\accepted{\today}
\submitjournal{ApJ}

\begin{document}

\title{Proper Motions of Sunspot's Umbral Dots at High Temporal and Spatial Resolution}

\correspondingauthor{Hadis Goodarzi}
\email{goodarzi@ipm.ir, hadisgoodarzy@yahoo.com}

\author[0000-0002-0786-7307]{Hadis Goodarzi}
\affil{School of Astronomy \\
Institute for Research in Fundamental Sciences (IPM) \\
P.O. Box 19395-5746, Tehran, Iran}

\affiliation{Research Institute for Astronomy and\\
 Astrophysics of Maragha (RIAAM), Maragha, Iran}

\author{Serge Koutchmy}
\affiliation{Institut d'Astrophysique de Paris UMR 7091 \\
CNRS and UPMC (Sorbonne University) \\
98 Bis Bd Arago, 75014, Paris, France}

\author{Ali Adjabshirizadeh}
\affiliation{Dept of Astrophysics, Faculty of Physics\\
Tabriz University, Tabriz, Iran}

\begin{abstract}

To deepen the analysis of the photometric properties of the umbra of a sunspot, we study proper motions of small features such as umbral dots (UDs) inside a
single sunspot observed by SOT of Hinode close to the disk center. We consider horizontal flows with high precision and details to study transient motion
behavior of UDs in short time intervals. Blue continuum images were first deconvolved with the PSF, such that the stray light is precisely removed and the original resolution is improved.
Several images were co-added to improve the S/N ratio keeping a reasonable temporal resolution and checking that the results are reproducible. The Fourier local correlation tracking (FLCT) technique is applied to the new corrected time sequence of
images and horizontal velocity maps were obtained both for the whole umbra (16$\arcsec$$\times$12$\arcsec$) and for a high resolution small region of
the umbra (3.5$\arcsec$$\times$3.5$\arcsec$) to study the smallest details of the velocity fields. We used two different Gaussian tracking windows (0.8
and 0.2 arcsec) which reveals two types of horizontal motions for umbral features. First, a global inner penumbra and peripheral umbra inward motion directed to the central parts is revealed as an overall proper motion of bright peripheral fine structures. Second, motions matching small cells inside the darkest parts of the umbra with apparent sink and source areas suggesting possible upflows and downflows appearing in different bright and dark locations without a definite answer regarding their brightness identification with a convective or a buoyant cell.

\end{abstract}


\section{Introduction} \label{sec:intro}

Study of the sunspots and their fine structures are one the most important aspects of solar activity physics \textbf{\citet{Adjabshirzadeh and Koutchmy1983}}. The velocity field inside a
sunspot presents a significant amount of details presumably related to the origin of the sunspot. Methods like Local correlation tracking technique applied to images obtained in the deep continuum allows to
determine the proper motions in horizontal directions. The analysis of spectral shifts of photospheric absorption lines permits to determine vertical or line of sight motions in higher levels where the lines are formed\citep[]{Ortiz et al.2010, Riethmuller et al., Bharti et al.2013} .

The existence of upflows and downflows in the umbra of a sunspot have been debated for several years
\citep[]{Riethmuller et al.,Deinzer1965,Denker and Verma2011}. For example \citet{Deinzer1965} concluded that convective motions should contribute
to radiation explaining the brightness of the umbra and the question of the field-free or of the magnetized nature of these motions is open \citep[]{Ortiz
et al.2010}. The value of the plasma beta inside the umbra is close to unity making both the mechanism of thermo- dynamical convection and the mechanism of
buoyancy as candidates for explaining up and down motions. Some evidences for upflows and downflows have been detected in sunspot structures as umbral dots
and light bridges \citep[]{Rouppe van der Voort et al.2010,Riethmuller et al.}. For instance, \citet{Watanabe et al.2012} detected clear upflows for
UDs using CRISP imaging spectropolarimeter data but could not detect systematic downflows associated with UDs. Also \citet{Riethmuller et al.} used
spectropolarization technique from Hinode/SP data by means of 2D inversion method to deduce line of sight velocity and managed to detect upflows and
downflows for peripheral UDs; note that this is above the deepest layers forming the continuum radiation that we study here.

Because of the small size of the umbral dots, and their varying size and velocity within a few minutes, studying motions of
UDs needs high spatial and temporal resolution. So we use the best available observations of the partial Sun free of turbulent Earth atmospheric effects
from the Solar Optical Telescope (SOT) onboard the Hinode spacecraft, after significantly enhancing the visibility and contrast by an optimum Maxlikelihood
deconvolution with the Point Spread Function (PSF) deduced in a preceding paper (\citet{Goodarzi et al.2015}, further designated paper I) and improving S/N ratio by co-adding several images. The Fourier local correlation tracking (FLCT) technique have been used with different time intervals, field of view and FWHM of the windowing Gaussian function for constructing velocity map of the sunspot UDs with different resolution to see any faint velocity changes \textbf{in} time. This technique was also applied to a granulation field of the nearby photosphere to demonstrate the correlation between converging (c) and diverging (d) proper motions in cell pattern with downflows of intergranular parts and upflows of bright granular parts, respectively. Besed on these these results, horizontal velocity proper motions maps can more convincingly also be used to consider vertical motion in the core of the sunspot by inferring sink (convergence) and source (divergence) area \citep[]{November1994,Sobotka et al.1999} making the same assumption of mass continuity in a stratified medium as for granulation.

\section{Observations} 

We used well focused images recorded with the broadband filter imager (BFI) of the solar optical telescope (SOT) onboard the Hinode spacecraft
\citep[]{Tsuneta et al.2008,Suematsu et al.2008}. Blue continuum images of the sunspot in NOAA AR 10944 that observed close to the disk center (heliocentric
angle of 4\degr) on March 1, 2007 were selected. The blue filter has a central wavelength at 4504.5\AA~ and a band width of 4\AA. Each pixel
corresponds to 0.05448$\arcsec$ on the surface of the Sun. Exposure time is 102 ms, the cadence is 6.4 seconds and the field of view is approximately
56$\arcsec$$\times$ 28 $\arcsec$. All of the images are first corrected for dark current, flat field and bad pixels that are removed using fg{\_}prep routine and then
co-aligned by means of two dimensional cross correlation algorithm involved in fg{\_}rigidalign routine (both of the codes are from Solar Soft library) and
then deconvolved with the PSF deduced in the previous paper \citep[]{Goodarzi et al.2015}. 90 successive blue continuum images taken between 00:19:51 and 00:29:20 with 6.4s time interval have been used to consider proper motions of fine structures inside sunspot umbra in high resolution, when adding images (at least 3 images averaging) to improve the signal to noise ratio.

\subsection{Fourier local correlation tracking technique} 

 In order to track proper motions of UDs, we used the Fourier local correlation tracking (FLCT) method \citep[]{Fisher and Welsch2008}. This technique
 permits to map a 2D velocity field that connects two images which are taken at two different times, such that this flow field provides the scalar field in
 the first image, the result having the most similarity to the second image \citep[]{Fisher and Welsch2008}. In this technique, it is possible to adjust a threshold value for intensities used to compute the flow field, so the algorithm will skip pixels with intensities lower than threshold
 value \citep[]{Fisher and Welsch2008}. Note that since the work on this paper, a more advanced algorithm appears, see \citet{Asensio Ramos et al.2017}
 where the possibility to compute the maps at 3 different levels is offered using much more extended computations.

 Up to now, the velocity field inside and around sunspots was described in several papers where ground- based images were used, see e.g. \citet{Molowny- Horas.1994} and \citet{Ji et al.2016}.
 Space data has the great advantage of being free from the turbulent earth atmosphere image effects which is very important for a correlation type analysis.
 In addition effects due to scattered light should be corrected. This is especially important when fine and low intensity structures are concerned such as umbral dots. We
 described in our previous paper \citep[]{Goodarzi et al.2015} how important it is to take into account the far wings of the point spread function (PSF) and only with this accountancy it is possible to remove stray light coming from far distances, indeed from the whole bright solar disk. We performed it by adding the Lorentzian function that has extended wings to the Gaussian function that describes the core of the PSF. We also use the limb of the sun for modeling stray light quite far from the solar disk instead of the Mercury or Venus disc of limited diameter; it permitted to go further and compute the coefficients of the Lorentzian and Gaussian functions more accurately.

\begin{figure*}
\resizebox{17.0cm}{!}{\includegraphics{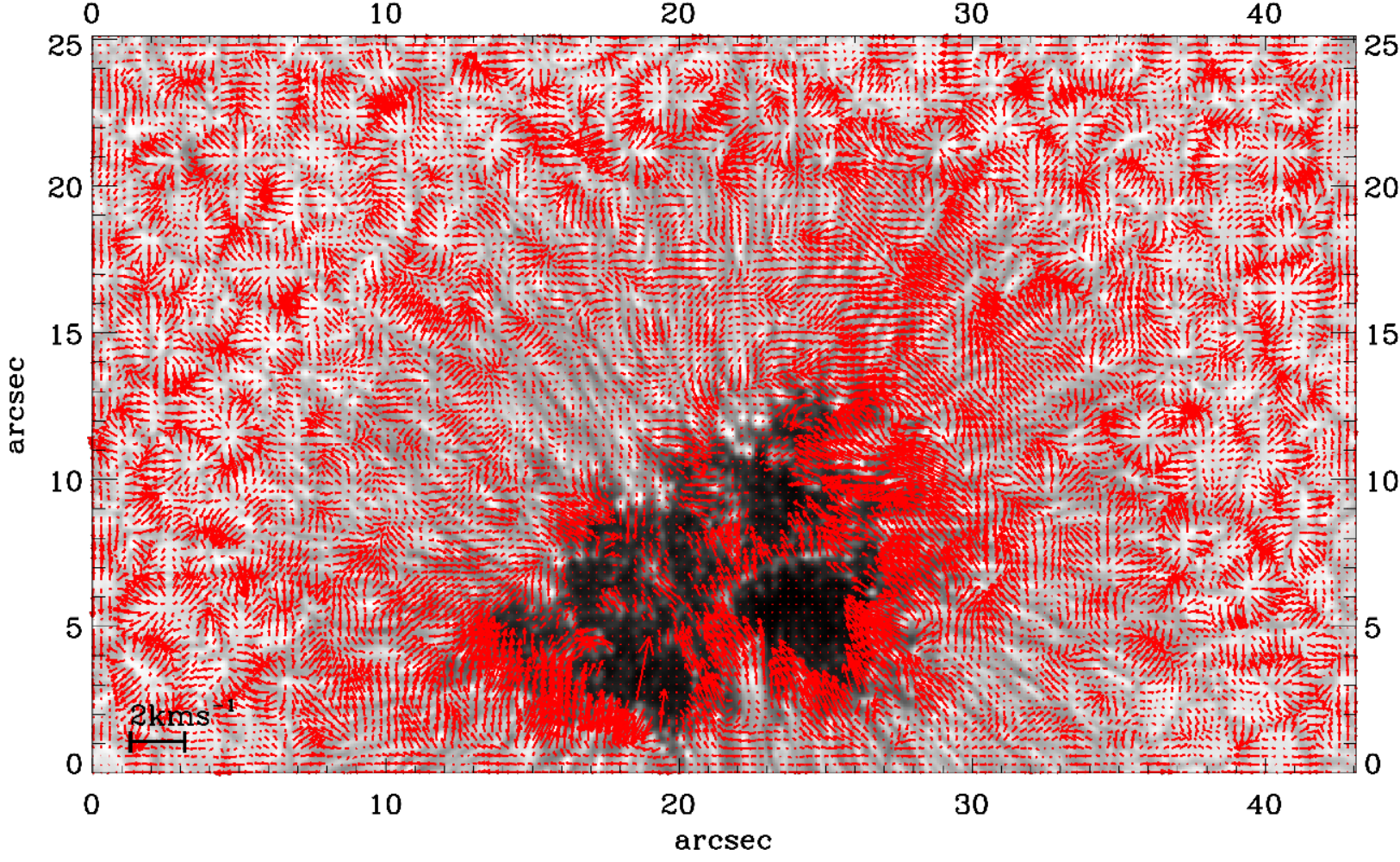}} \hfill \caption{Proper motion (horizontal Velocity component) map in the sunspot observed on March 1th,
2007 as deduced from using the FLCT technique (FWHM of Gaussian windowing function is 0.8$\arcsec$). The scale for velocity is inserted at the bottom left of the figure and corresponds to 2Km/s.
Note the systematic inward proper motion of the inner penumbral features and dots and better use a magnified image from the digital variant of the paper.}
\label{fig:1}
\end{figure*}

 After deducing the PSF precisely, a deconvolution procedure was performed using the iteration method \citep[]{Goodarzi et al.2015,Goodarzi et al.2016}. Correction with this method leads to a higher contrast and a more deeper evaluation of the images that are now free of stray light. In order to improve the signal to noise ratio, several images (at least 3 images, up to 10 images) were co-added and then the FLCT technique was applied to successive average images to deduce the velocity
 fields. We performed our analysis of sunspot umbra in 2 parts: an overall proper motion  mapping of the whole umbra (including the inward motion of peripheral umbral dots) see fig.~\ref{fig:1}, and a mapping of a small selected region, which represents a part of the sunspot umbra, to see the ultimate details of proper motions.

\begin{figure*}
\resizebox{17.0cm}{!}{\includegraphics{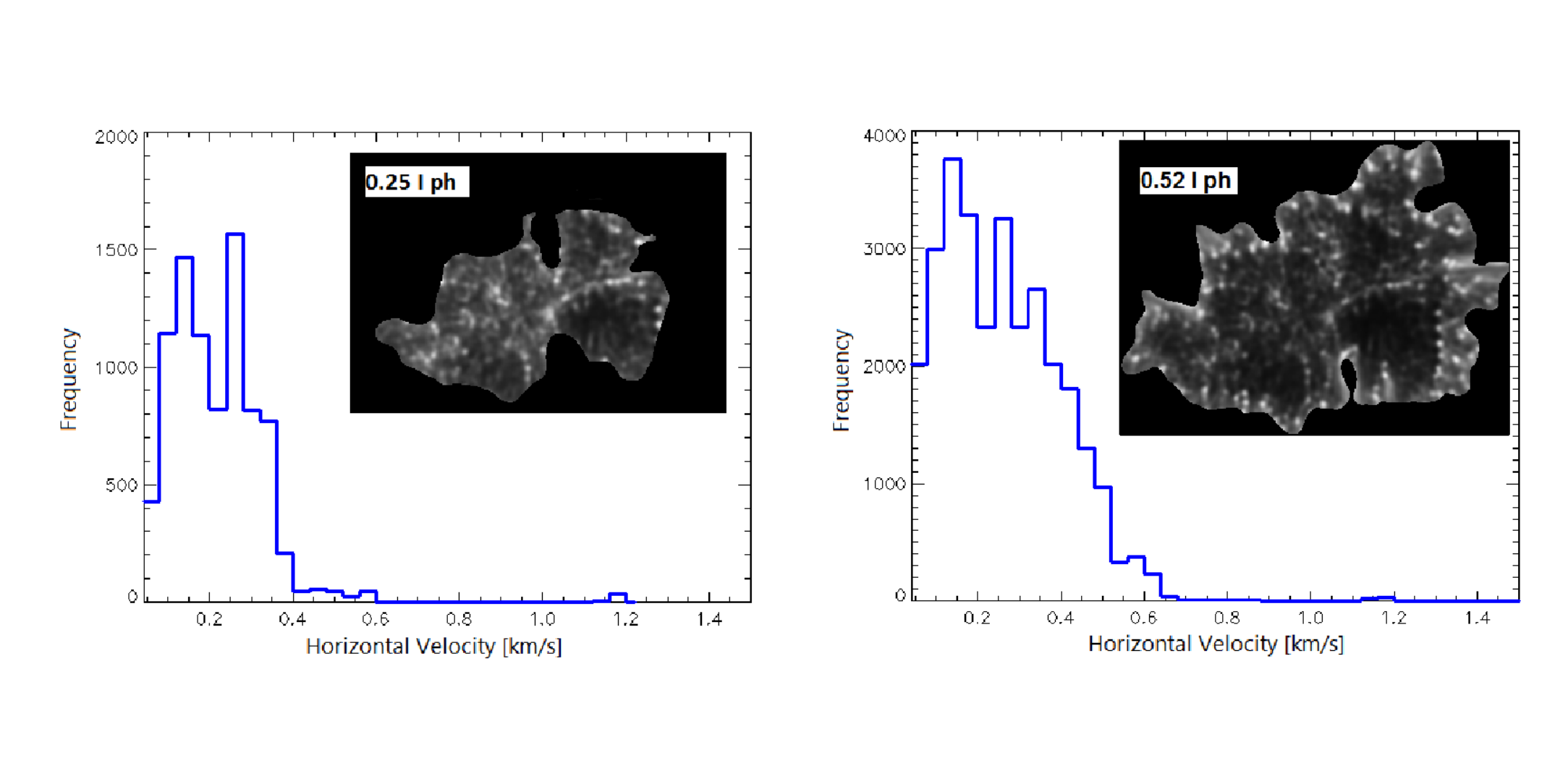}} \hfill
\caption{Histogram of horizontal velocities (in km/s) inside the sunspot for two different thresholds, 0.25 and 0.52 of the average photosphere intensity
revealing velocity values for only the central region of the umbra (left), and of the whole umbra (right) including the peripheral UDs.}
\label{fig:2}
\end{figure*}

\section{Overall proper motion of the whole umbra}

To see the overall proper motions of umbral features we applied the FLCT technique on the images of the sunspot observed on 1th March 2007 after
deconvolution and co-adding images to improve the signal to noise ratio. This part is specially done to demonstrate that the result of running the FLCT code is
altogether in agreement with the results of other researches (for example \citet{Molowny- Horas.1994}; \citet{Sobotka et al.2008}) but in the details, there is a lot of new features. It is also in agreement with the impression given after watching the movie of our deconvolved images, see our paper I.

In order to analyze the overall proper motions of umbral features and UDs, we selected the entire field of view for running the FLCT code with a rather
large FWHM of the windowing Gaussian function (0.8 arcsec) used for the mapping. Fig.~\ref{fig:1} shows the velocity map of the whole sunspot region
obtained using deconvolved images and averaging 3 successive images made with a 6.4 sec interval between 2 successive images (19 sec averaging) and shifting
the averaging by 6 minutes after again another 3 successive images were averaged (without overlapping) to see the motion of the umbral and the penumbral
features during this 6 minutes interval between two average images. Also we imposed a threshold value equal to approximately 15 percent of the average photospheric intensity for using the FLCT code to select just the behavior of the very bright features, so the code skips pixels with lower intensities dominated by the noise. For the corrected intensity values that we used, see our paper II \citet{Goodarzi et al.2016}. Inward motions of the peripheral UDs \textbf{are} very well apparent in this velocity map. Note also in the granular region of the image, the downflow in the dark intergranular region and upflows in the bright parts clearly seen all around the sunspot.

It can be seen from Fig.~\ref{fig:1} (better use a magnified image from the digital variant of the paper), our result of running the FLCT code  on deeply deconvolved images is in overall agreement with \citet{Molowny- Horas.1994}, \citet{Ji et al.2016}
and \citet{Sobotka et al.2008}.
As Fig.~\ref{fig:1} suggests, peripheral UDs show higher horizontal velocities in comparizon with central UDs. To compare this quantitively, two Histograms
of the velocity values have been prepared: one of them includes both peripheral and central UDs and one of them includes only velocities of the central part
of the umbra. This is performed using two different threshold values, corresponding to approximately 50 and 25 percent of the average frontier intensities. As it can be seen from the histograms in Fig.~\ref{fig:2} and the velocity map of Fig.~\ref{fig:1}, when peripheral umbral dots are included, higher value of the
horizontal velocities (between 400 and 700 m/s) are more likely to exist and it means that features faster than 400 m/s are usually found in the peripheral
part of the umbra where magnetic field is weaker than inside the core and is more horizontal that is consistent with the results of \citet{Sobotka and Puschmann2009} and also \citet{Watanabe et al.2009}.

\section{Inwards motion of peripheral UDs}
\begin{figure}
\resizebox{8.0cm}{!}{\includegraphics{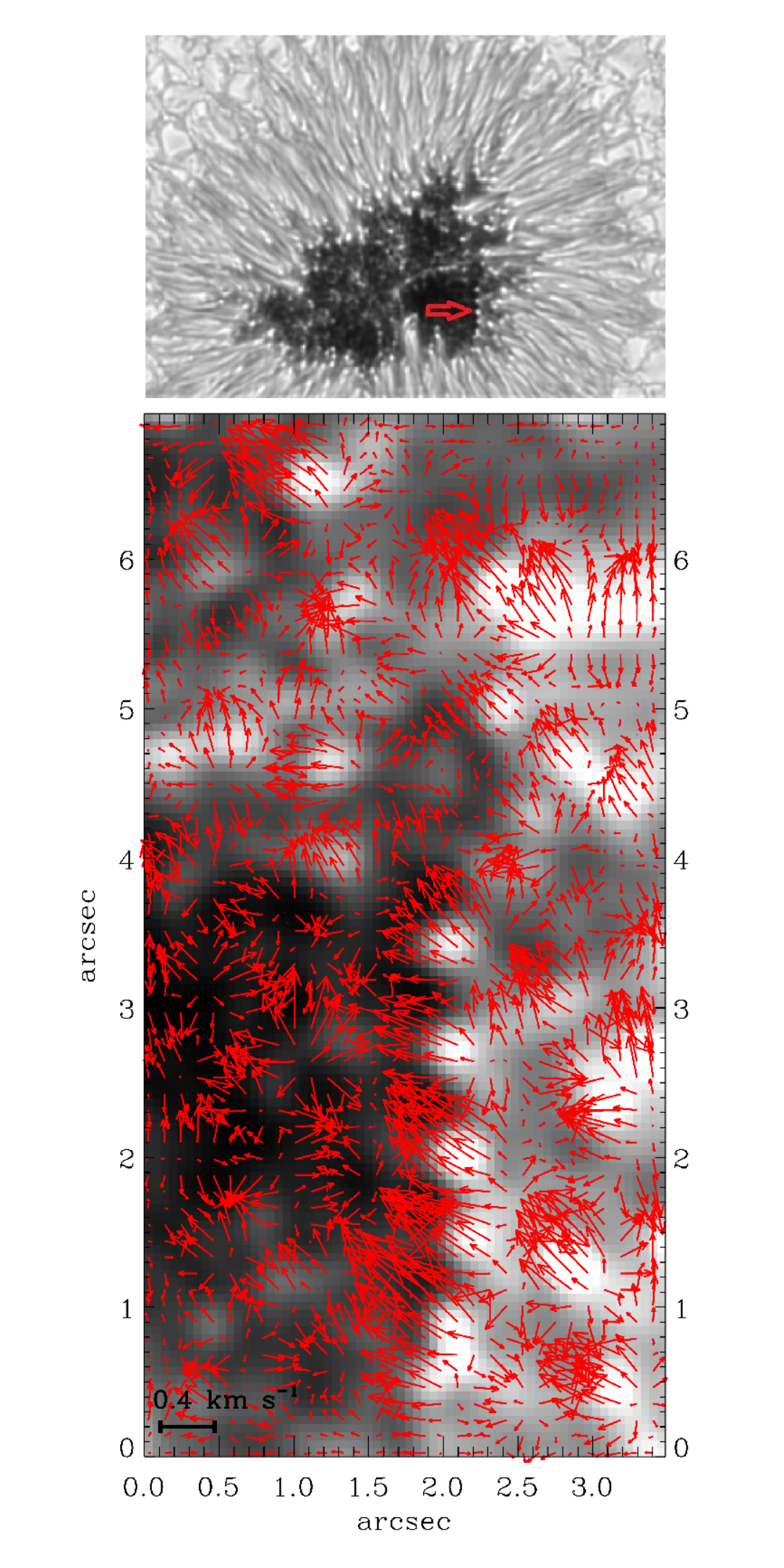}} \hfill
\caption{Top: Sunspot image observed on March, 1th 2007 after correction with the deconvolution procedure. Red arrow show the part of the umbra that is used to consider in details inwards motion of the peripheral UDs. Bottom : Velocity map for a partial frame of the umbra including peripheral UDs (indicated with red arrow at the top of image). \textbf{FWHM of Gaussian windowing function is 0.2$\arcsec$}. Peripheral UDs show well inwards motion to the umbra but more details are revealed.}
\label{fig:3}
\end{figure}

In velocity maps derived with different methods, peripheral UDs show the inwards motion towards the umbra \citep[]{Hamedivafa2015}. These motions
can be clearly seen from  Fig.~\ref{fig:1}, but to consider them with more details and with a higher resolution, we made velocity map for a small part of
the umbra-penumbral boarder (also called the periphery of the umbra) and using a much smaller FWHM of the windowing Gaussian function (0.2 arcsec). This is made achievable since that using accurately deconvolved images free of scattered light \citep[]{Goodarzi et al.2015} and after improving the S/N ratio at an optimum level to warranty the reproducibility of the results, the most significant intensity distributions (see the next part) are used.
The selected part is shown with red arrow in top part of Fig.~\ref{fig:3}, using a deconvolved sunspot full image observed on 1th of March 2007; this field includes several bright peripheral UDs. Bottom part of fig.~\ref{fig:3} shows the result of using the FLCT technique for this region (we have 6 consecutive images with 6.4s cadence, averaging every 3 images leads to two average image and fig.~\ref{fig:3} is the result of a FLCt run from comparison of these two average images). As it can be seen peripheral umbral dots have inwards motion to the umbra but some more details are also revealed like the small diverging and/or converging flows that accompany with this inwards motions.

\section{High resolution velocity maps} \label{sec:HR velocity map}

We presented in  Fig.~\ref{fig:1} the global proper motion map of the sunspot field including umbra, penumbra and also the granulation part of the surrounding photosphere by imposing a
threshold value for using the FLCT code to follow only the very bright features while using a large FWHM of the windowing Gaussian function (0.8 arcsec).

\begin{figure*}
\resizebox{17.0cm}{!}{\includegraphics{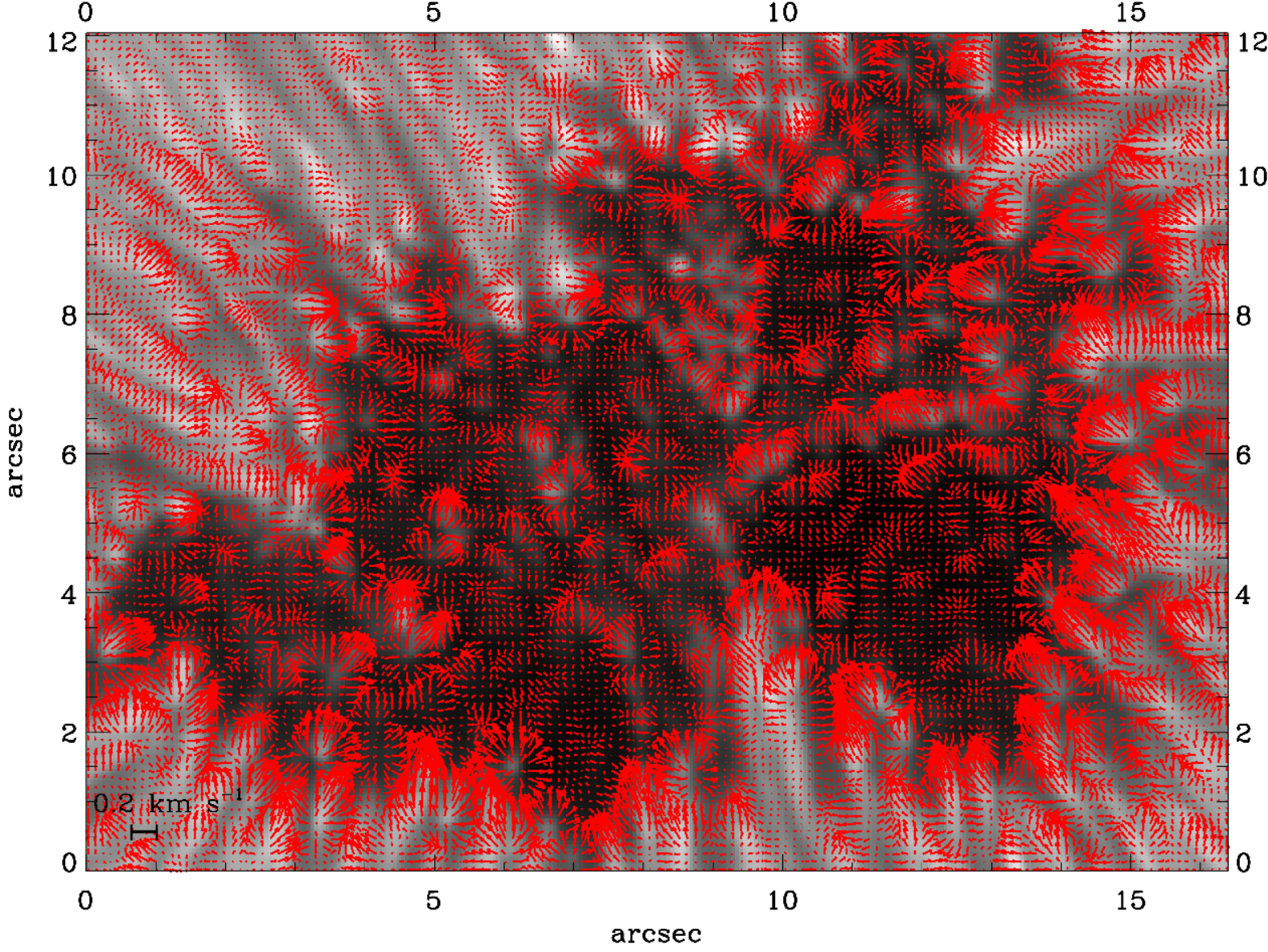}} \hfill
\caption{ Detailed horizontal velocity map including the darkest umbral parts of the sunspot observed on March 1th, after significantly
improving the S/N ratio (see Section~\ref{sec:HR velocity map}). FWHM of the Gaussian windowing function is 0.2$\arcsec$. The scale inserted at the bottom left corresponds to 200 m/s.}
\label{fig:4}
\end{figure*}

Now, to see the details in full resolution, a smaller FWHM of the windowing Gaussian function has been used for the whole umbra without any threshold value. Because we take into account all pixels including pixels corresponding to low intensities (especially in the darkest parts of the umbra), we have to significantly improve the S/N ratio by co-adding more images (10 images) to reduce the noise that would contribute in deriving velocity maps. Fig.~\ref{fig:4} shows the proper motion map of the sunspot umbra which is obtained after deconvolution of 90 images (after making evaluation tests to optimize the procedure), an averaging of 10 successive images with a 6.4sec time interval between them (1 min averaging) and shifting the averaging procedure to the next 10 images with a 3 images overlapping and computing velocity field from these average images. Averaging procedure continue up to 90 images that leaded to 11 resulting velocity maps. Fig.~\ref{fig:4} is the average of these 11 velocity maps corresponding to a duration of 9.6 min.

From Fig.~\ref{fig:4} it can be seen that [i] Peripheral umbral dots show the inwards motion to the umbra with additional
details revealing converging and diverging flows; the converging and diverging behavior happen more clearly in the endpoint of the penumbral filaments or peripheral UDs with higher S/N ratio in comparison with central UDs, which is consistent with \citet{Riethmuller et al.}. [ii] In the penumbral
part of the image (top left) we see flows that are perpendicular to the bright filament and that move across the bright filament going to the dark filament.
[iii] In the umbra, there is an evidence of both converging and diverging horizontal motions everywhere.

\begin{figure}
\resizebox{8.4cm}{!}{\includegraphics{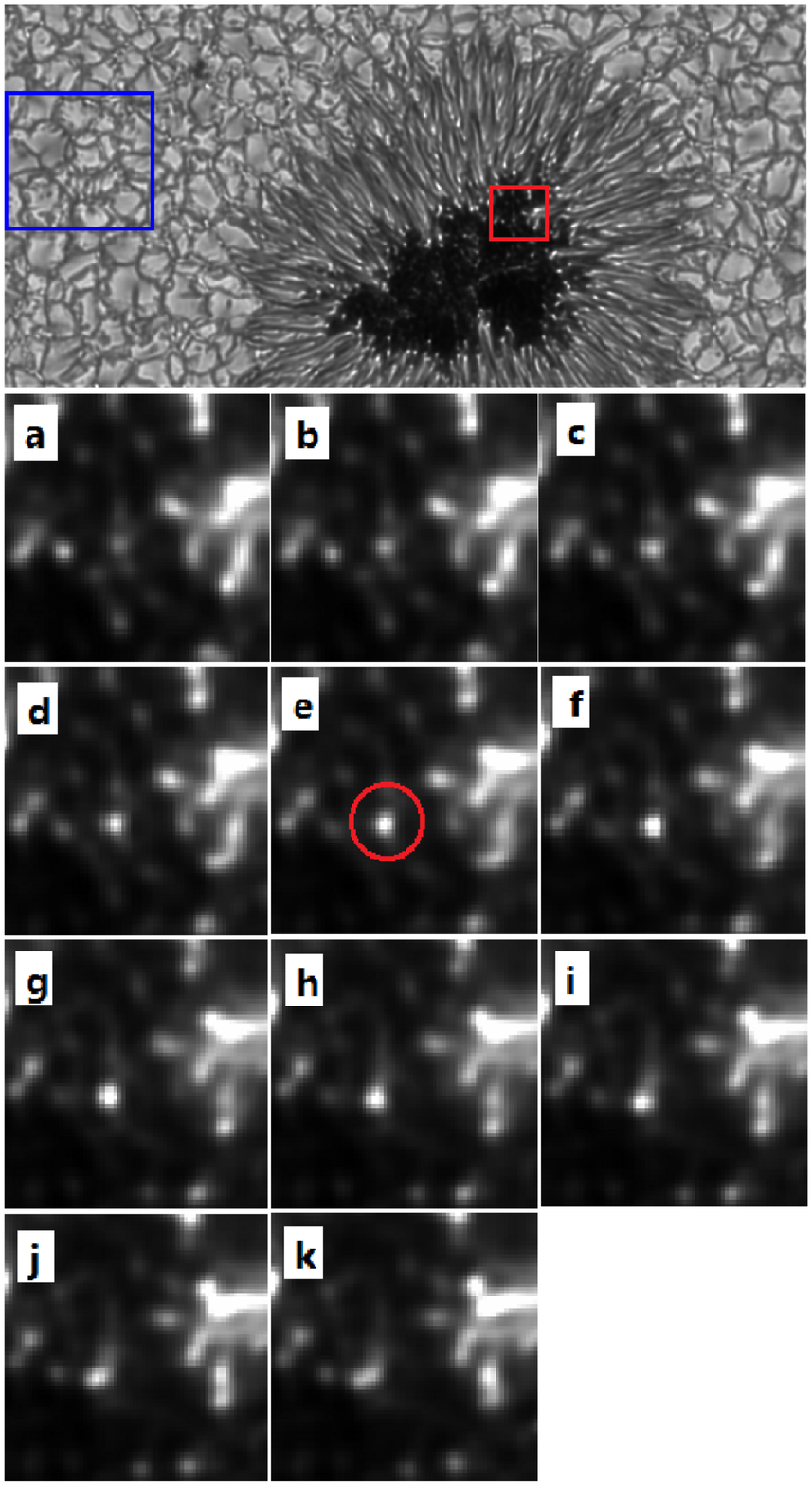}} \hfill
\caption{Top: Deconvolved Sunspot image with the blue box showing a selected field of view for making proper motions maps of granulation part (Fig.~\ref{fig:6}) and red box showing a selected field of view for computing horizontal flow field during 9.6 minutes in small part of the umbra as illustrated in
Fig.~\ref{fig:8}
Mosaic of the averaged images of a selected part of the sunspot (red box at the top of image) evolving during 9.6 minutes with the isolated bright UD of short lifetime shown in window "e".}
\label{fig:5}
\end{figure}

\begin{figure*}
\resizebox{15.0cm}{!}{\includegraphics{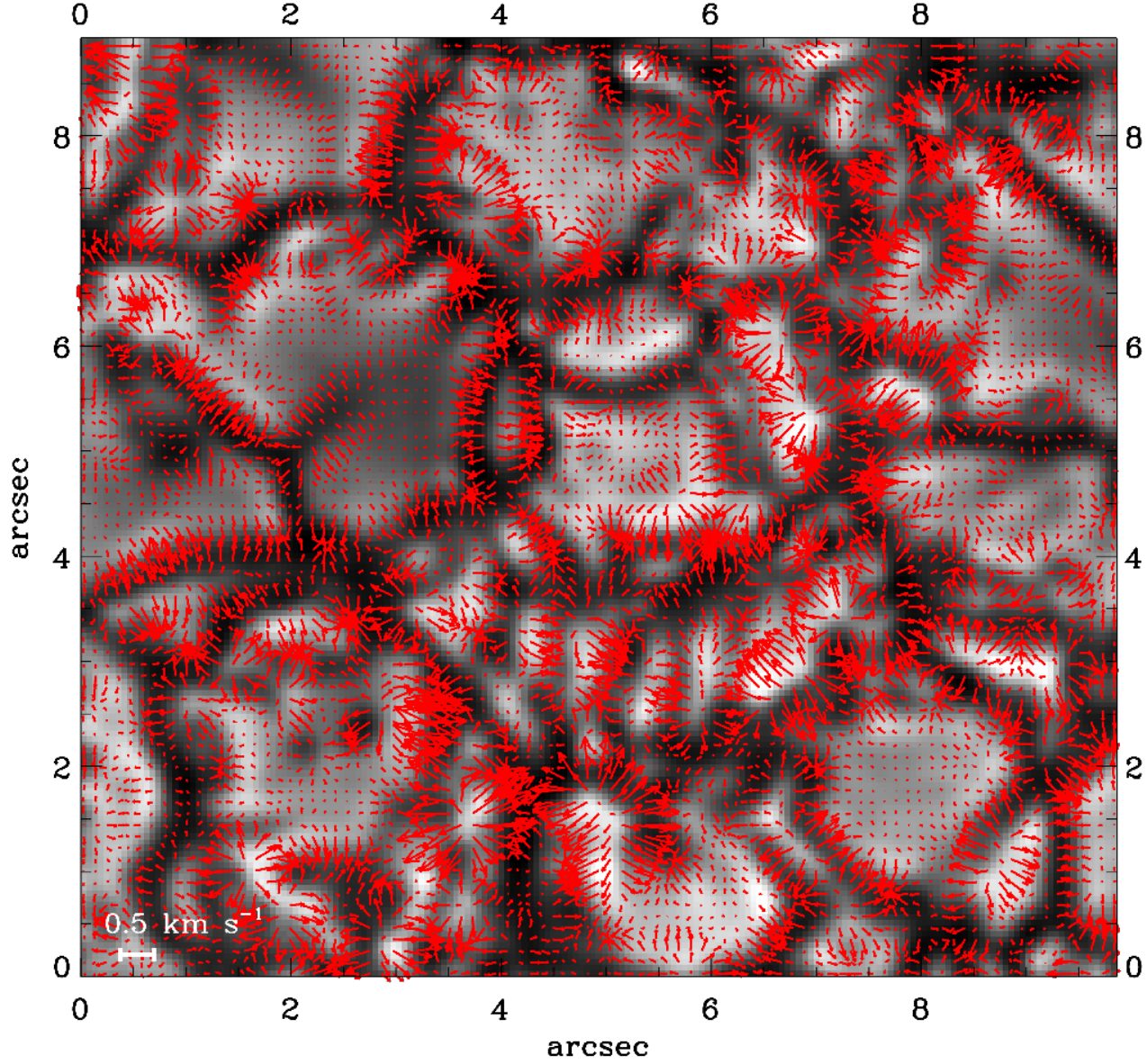}} \hfill
\caption{Horizontal velocity field computed for selected granulation part near the sunspot shown in the blue box in top part of Fig.~\ref{fig:5}. FWHM of Gaussian windowing function is 0.2$\arcsec$.}
\label{fig:6}
\end{figure*}

\section{INTERPRETING UPFLOWS AND DOWNFLOWS FROM CONVERGING AND DIVERGING HORIZONTAL MOTIONS}
FLCT technique has being used to deduce horizontal motions of different solar structures. However some upflows and downflows can be inferred from the converging and diverging arrows distribution, as it has been done for the granulation starting from the works on SOUP data, see \citet{Simon et al. 1989}. In order to examine this technique to presumably infer vertical motions, we applied it to a granulation part of the adjacent photosphere surrounding the sunspot, indicated with a blue rectangle in top part of figure Fig.~\ref{fig:5}. This field was submitted to exactly the same processing (cadence, corrections, deconvolution and summing) that what was done for the UDs field (see after). We assume as usual that bright granules are convective cells. Accordingly, they correspond to upward motions and conversely, intergranular space corresponds to downward motions. We also assume that for the small scales corresponding to our analysis, 5 min oscillations do not perturb the results.
Every 10 images with 6.4s cadence have been added with 3 images overlapping and the proper motion between the first two averege images (45s time interval) has been computed and depicted in Fig.~\ref{fig:6}. Gaussian width used for weighting images equals to 0.2$\arcsec$. In this velocity map, we see conveging flows in intergranular parts and diverging flows also appear in several bright points of granular parts which corresponds to downflows and upflows respectively.

We notice that these apparent flows could be also due to brightness changes of coherent but unknown nature, however it is not legitimate to ignore the presence of downflows and upflows contributing to these brightness changes in order to take into account the convective phenomena that are today well explained by impressive theoretical simulations, see for e.g. \citep[]{Stein and Nordlund1998}. The good correlation of lanes with downflow in intergranular parts and diverging pattern as cells with upflow in bright granular parts shows that the FLCT techniques permit a valuable evaluation of vertical motions using the parameters of our analysis.

\begin{figure*}
\resizebox{17.0cm}{!}{\includegraphics{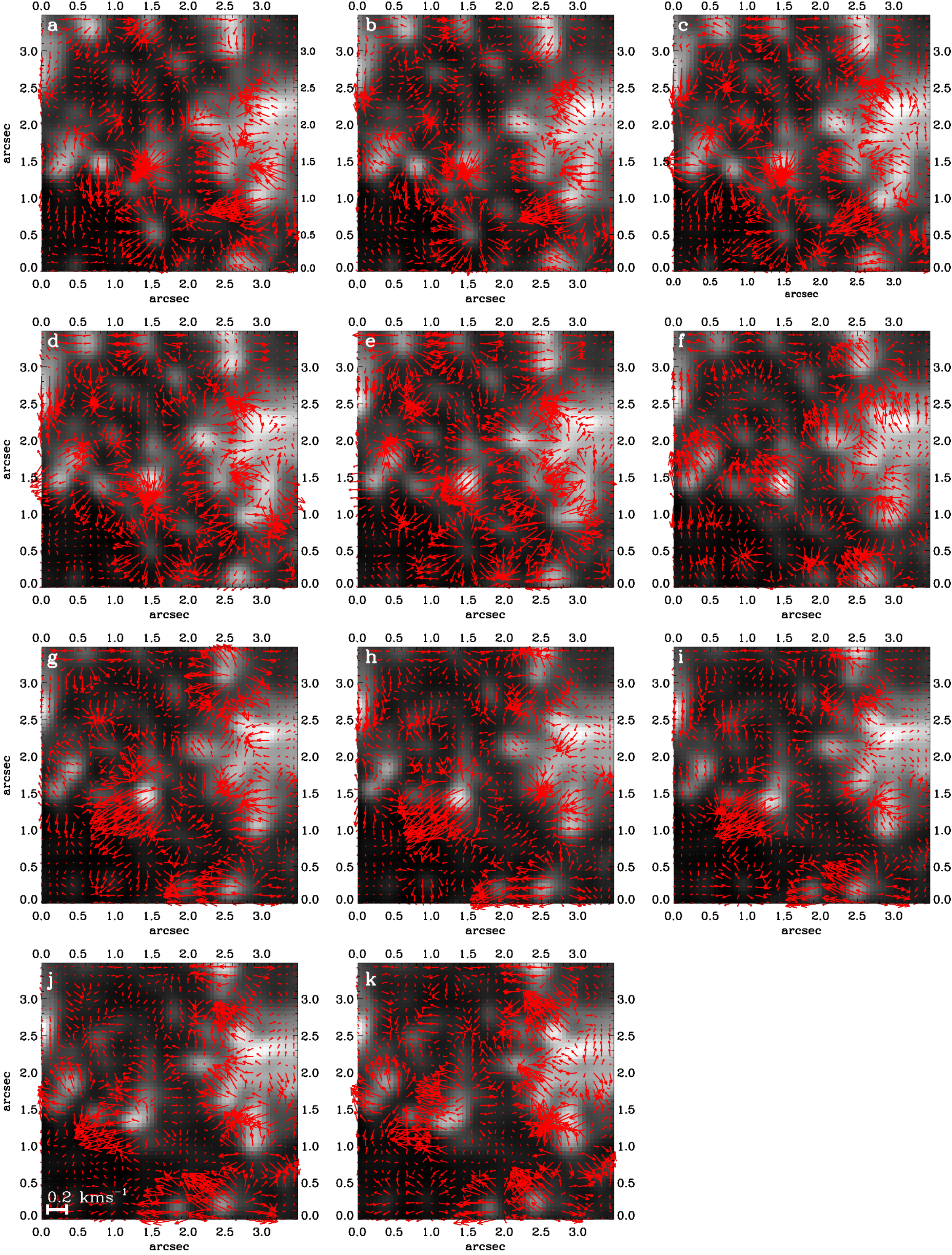}} \hfill
\caption{Velocity field computed from FLCT technique for small field of view (red box in top part of Fig.~\ref{fig:5}) and for short time intervals (45s). FWHM of Gaussian windowing function is 0.2$\arcsec$. Red arrows show the direction of velocity for each point and its length shows the amplitude of the horizontal component of the velocity compared
to the scale inserted in part j (bottom, left) which corresponds to 200m/s.}
\label{fig:7}
\end{figure*}
\begin{figure}
\resizebox{10.4cm}{!}{\includegraphics{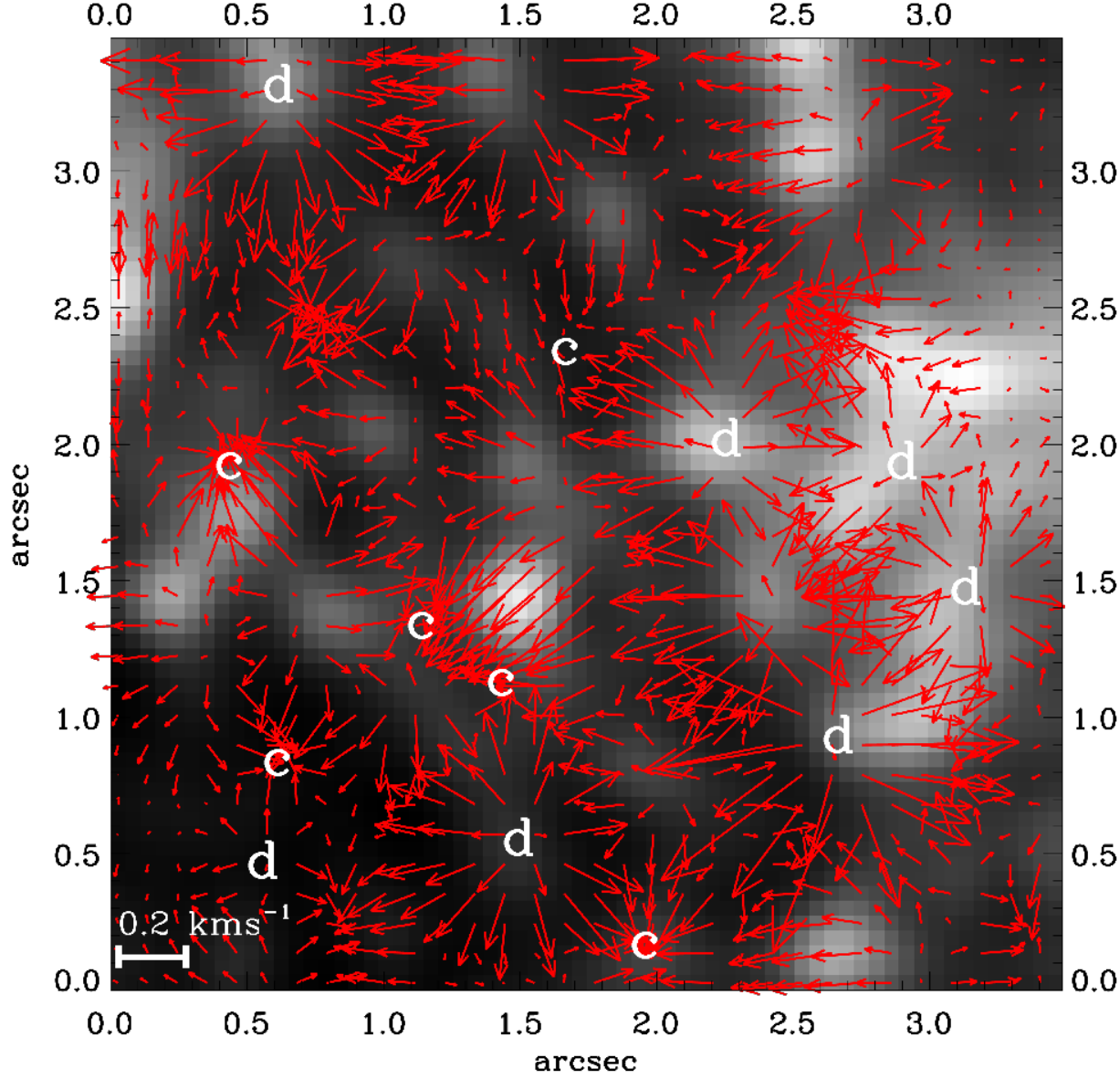}} \hfill
\caption{Velocity field computed from FLCT technique for small field of view and for short time intervals (45s) with labels showing \textbf{converging (c)} and
diverging (D) flows. FWHM of Gaussian windowing function is 0.2$\arcsec$.}
\label{fig:8}
\end{figure}

\section{Proper motions of UMBRAL DOTS AT ULTIMATE RESOLUTION }
In velocity maps of Fig.~\ref{fig:3} and Fig.~\ref{fig:4} it was clear that using the high quality deconvolved and averaged images reveals more details of
proper motions and some diverging and converging area appear in both the bright and the dark regions. Also some of the motions change their orientation in
short time intervals (less than 1 min) that might be ignored in velocity maps derived from images with longer time interval (Fig.~\ref{fig:4} that
is an average over \textbf{9.6} min). To see the maximum of details and the tiny variations of proper motions using the ultimate resolution from this analysis, a set of new velocity maps was evaluated on a small field of view (3.5$\arcsec$$\times$3.5$\arcsec$) and short time interval (45s) after corrections and improving the signal to noise
ratio.

For this purpose, we used 90 successive blue filter images with 6.4s cadence after deconvolution that cover approximately \textbf{9.6}min. Each set of 10
images was co-added with 3 images overlapping in order to improve the signal to noise ratio and also to preserve the continuity between images, so 12 average
images were deduced with approximately 45s time interval between them.  Then a small field of view that includes a part of the inner penumbra and some peripheral UDs is selected from the sunspot image to
see the tiny velocity field attributed to umbral features. This selected part (that is shown with red box in top part of Fig.~\ref{fig:5}) is specially chosen because of the very typical case of a well observed isolated bright UD (see the red cicle in part "e" of Fig.~\ref{fig:5}), which is approximately roundish and evolutionally appealing. Gaussian width used for weighting images equals to 0.2$\arcsec$ that is almost the smallest reasonable width for a tracking window keeping a reasonable signal/noise ratio.

From the 12 average images, 11 velocity map are prepared and a mosaic of them is depicted in Fig.~\ref{fig:7} and they are also shown in a short on-line video (made from 11 frames) which is available on the electronic version of the paper.

 Bottom part of Fig.~\ref{fig:5} shows the time brightness evolution of the selected small part of the umbra with a cadence of 45s (red box in top part of Fig.~\ref{fig:5}) including the interesting case of a well isolated umbral dot indicated with a red circle in "e". This umbral dot first appear faintly (a) and then become brighter (f, g) and finally fade after approximately 9.6 minutes when its direction of motion changing during time which can be seen better with the high resolution in the velocity maps of Fig.~\ref{fig:7} (or in the movie).

 From the velocity maps (Fig.~\ref{fig:7} and the movie) two kind of motions can be distinguished at short time interval (45s) and for small field of view with a much improved resolution: i/ Proper motions toward a preferred direction, for example position (x,y)=(0.8, 1.4) in part"a" of Fig.~\ref{fig:7}. Some of these motions can change slightly their trajectory, by tracking the case of isolated umbral dot (position (x,y)=(1.5, 1.5) in part "a" of Fig.~\ref{fig:7}.) it can be seen that the direction of its horizontal motion changes during a small time interval but there is a dominant movement toward the umbra. As noted before, the most dominant horizontal motion of the bright points (like the one marked with red circle in bottom part of Fig.~\ref{fig:5}) is toward the umbra, but some small scale turbulent behavior may be seen also. ii/ Some flows in the form of converging (c) and diverging (d) cells that could correspond to possible inward and outward motions as deduced from considerations of continuity. Outward motions (source area, where red arrows tend to show up further from each other, see for example Fig.~\ref{fig:8}) are sometimes seen near some bright points, while there are cases of outward motions in dark area. The center of several diverging velocity pattern ("d" cells) are rather side-on of UDs maximum intensity brightness.

 Since the occurrence of vertical flows preferentially in deep photospheric layer were considered in theoretical works e.g. \citet{Riethmuller et al.}, images taken in the blue continuum that we used correspond the deepest layer of the photosphere
 and they show suitable candidates for revealing these flows. Source field associated with bright UDs are not always overlapping the maximum brightness of UDs
 and some displacement can be seen as also appearing in other works \citep[]{Ortiz et al.2010}. There are also some sink areas (or downflows) that mostly
 appear at the periphery of UDs or in the border region between bright features and this is rather consistent with \citet{Riethmuller et al.} works
 which found downflows in 200 to 500 km distance from UD center which are concentrated on one or two sides of UDs. Since we could not detect any dark lane
 inside UDs, it is not possible to follow their endpoints looking downflows as \citet{Schussler and Vogler2008} predicted from their most detailed
 magnetoconvective simulations but we can say that they mostly appear only at the side of the bright features, see also our paper II. In addition our observation show fast changes, over 1 min of time, which is not seen in published magneto-convection simulations.

\section{conclusion}
We detected proper motion of umbral fine features from corrected (restored by deconvolution of the measured PSF) blue continuum images with different time intervals and spatial scales show new features in the small scale velocity maps. The most dominant motion of fine structures is toward the umbra but when higher resolution is considered, more details can be seen such as the complex cells of converging and diverging flows which could be the sign of vertical motions of plasma. These possible upflows usually appear near bright features, while downflow evidences tend to exist at the periphery of bright features or in the border between them.

 We could not see any dark lane inside of UDs predicted by numerical simulation
\citep[]{Schussler and Vogler2008,Bharti et al.2010}, but their prediction of upflows and downflows in and around UDs can still be consistent with our
observational findings, taking into account [i] the limited resolution of our observations and [ii] the limitation of the interpretation of our flow map of
horizontal velocities in term of upflows (source) and downflows (sink); [iii] the approximations used in the simulations.

Our results are complementary to the results obtained from the analysis of spectral
line shifts in the regions above the continuum layers \citep[]{Ortiz et al.2010,Riethmuller et al.}. Furthermore, we provide evidences for
systematic fast motions inside the core of the probed sunspot.


When a time sequence is considered, the temporal variation of the velocity reveals a sort of turbulent behavior in
the trajectory of the bright UDs that was analyzed in details. Horizontal proper motions that are not reproduced in the recent numerical simulations that we know are evidenced. The example of a dark region of the core showing some diverging cells appearing temporary is pointed out. Accordingly, it is not inappropriate to remind about the umbral oscillatory convection theoretically studied in \citep[]{Weiss et al.1990} in the frame of the magneto- convection inside a sunspot. The analysis of a much longer time sequence taken at different formation heights is desirable to go in this direction. Given the improved S/N ratio (Fig.~\ref{fig:4} and Fig.~\ref{fig:7}), a detailed and complete map of the entire umbra including the darkest regions has been obtained. We note that when we point to the systematic motions towards the umbra, it does not necessarily mean a direct movement and some turbulent behavior also exist that is
revealed by tracking a sequence of high resolution velocity maps (see also the movie of paper I).


\acknowledgments
Hinode is a Japanese mission developed and launched by ISAS/JAXA, collaborating with NAOJ as a domestic partner, NASA and STFC (UK) as international
partners. Scientific operation of the Hinode mission is conducted by the Hinode science team organized at ISAS/JAXA. This team mainly consists of scientists
from institutes in the partner countries. Support for the post-launch operation is provided by JAXA andNAOJ (Japan), STFC (U.K.), NASA, ESA, and NSC
(Norway). We thank Habib Khosroshahi for providing a critical reading of the paper and our reviewer for meaningful remarks. This work has been supported financially by Research Institute for Astronomy and Astrophysics of Maragha (RIAAM) under research project No.
1/5237-69.



\end{document}